\documentclass{article}
\usepackage{graphicx} 
\usepackage[utf8]{inputenc}
\usepackage{authblk}
\usepackage{caption}
\usepackage{hyperref}
\usepackage{mathtools, nccmath}
\usepackage{amssymb}
\usepackage{geometry}
\title{Simple few-shot method for spectrally resolving the wavefront of an ultrashort laser pulse}
\author[1,3,*]{Slava Smartsev}
\author[2,3]{Aaron Liberman}
\author[1]{Igor A. Andriyash}
\author[1]{Antoine Cavagna}
\author[1]{Alessandro Flacco}
\author[1]{Camilla Giaccaglia}
\author[1]{Jaismeen Kaur}
\author[1]{Jos\'ephine Monzac}
\author[2]{Sheroy Tata}
\author[1]{Aline Vernier}
\author[2]{Victor Malka}
\author[1]{Rodrigo Lopez-Martens}
\author[1]{J\'er\^ome Faure}

\affil[1]{Laboratoire d’Optique Appliquée, ENSTA Paris, CNRS, Ecole Polytechnique, Institut Polytechnique de Paris, 828 Bd. des Maréchaux, 91120 Palaiseau, France}
\affil[2]{Department of Physics of Complex Systems, Weizmann Institute of Science, 234 Herzl St., Rehovot 7610001, Israel}
\affil[3]{The authors contributed equally to this work}

\affil[*]{smartslava[at]gmail[dot]com}

\date{}

\usepackage{lineno}
\usepackage{float}

\begin{document}

\maketitle

\begin{abstract}
We present a novel and straightforward approach for the spatio-spectral characterization of ultrashort pulses. This minimally intrusive method relies on placing a mask with specially arranged pinholes in the beam path before the focusing optic and retrieving the spectrally-resolved laser wavefront from the speckle pattern produced at focus. We test the efficacy of this new method by accurately retrieving chromatic aberrations, such as pulse front tilt, pulse front curvature, and higher-order aberrations introduced by a spherical lens. The simplicity and scalability of this method, combined with its compatibility with single-shot operation, make it a promising candidate to become a new standard diagnostic tool in high-intensity laser facilities.
\end{abstract}

\section{Introduction}
Ultrashort lasers, with femtosecond pulse durations, are indispensable tools in medicine, industry, and science. The advent of chirped pulse amplification (CPA) \cite{STRICKLAND_OC_1985} enabled such pulses to be amplified to peak powers of terawatts and even petawatts \cite{Danson_HPLSE_2019}. Intense, short laser pulses have opened possibilities to explore laser-matter interactions in the relativistic regime \cite{mourou-rmp06}, and their applications as compact laser-plasma accelerators \cite{Esaray_RMP_2009}.

Ultrashort lasers are broadband and the spectral phase can impact the pulse's duration and shape in the temporal domain. Accurate and precise measurement of the spectral phase is the core of temporal metrology of ultrashort pulses \cite{Walmsley_09}. Typical measurements assume the spectral phase is not spatially varying.

High-intensity lasers, however, operate with large beams for which the spatio-spectral phase can be very important. Spatially non-uniform dispersion effects accumulated during amplification, propagation, and focusing can lead to different spectral components of the beam ending up with different wavefronts. These wavefronts determine how and where each color is focused, ultimately shaping the full multi-color spatiotemporal intensity at the focus. These and related effects are usually referred to as spatiotemporal couplings (STCs) \cite{Akturk_2010}.

The most commonly encountered STCs are pulse-front tilt (PFT), which can be caused by the misalignment of the gratings in the compressor, and pulse-front curvature (PFC), induced by chromatic lenses \cite{Bor_89}. In most cases, STCs are undesirable because they increase pulse duration and reduce peak intensity and contrast at the focus \cite{Bourassin_OE_2011}. In some cases, however, STCs can be exploited in a controlled and intricate way to manipulate the dynamics of intense pulses in the focal region \cite{Sainte_Marie_Optika_2017,Froula_NPH_2018}. Control over the velocity with which energy is deposited along the focal region, through the tailoring of STCs, paves the way toward a new generation of laser-driven particle accelerators \cite{Debus_2019_PRX,Caizergues_NP_2020,Palastro_PRL_2020} and X-ray sources \cite{Kabacinski_2023}. Techniques for accurate and straightforward measurement of STCs are necessary to mitigate unwanted couplings and enable the utilization of STCs as a critical degree of freedom in an experiment.

The development of spatiotemporal metrology is well summarized in several reviews \cite{Akturk_2010,Dorrer_2019,Jolly_IOP_2020}.
The main methods of STC characterization of ultrashort pulses are listed below.

(1) There are methods that utilize spectrally-resolved wavefront measurements based on Shack-Hartmann sensors. HAMSTER, for instance, uses an acousto-optic programmable dispersive filter to isolate spectral components and then a Shack-Hartmann sensor to reconstruct each component's wavefront \cite{Cousin_OL_2012}. Other similar techniques employ optical filtering to narrow the pulse's spectral content \cite{Kim_OE_2021,Weise_2023}.

(2) A number of methods are based on spatially-resolved Fourier-transform spectroscopy (FTS). A self-referenced version of FTS, TERMITES \cite{Miranda_OL_2014,Pariente_NatPhot_2016}, uses a spatially filtered copy of the beam as a clean reference for interference in the near field (NF). INSIGHT \cite{Borot_OE_2018} uses a similar approach, but the interference is observed in the far field, and the Gerchberg–Saxton (GS) iterative algorithm \cite{GS_1972} is used to retrieve the spatially resolved spectral information of the beam.

(3) Some methods use hyperspectral imaging techniques for STC measurements. One example is broadband ptychography which is based on coherent diffraction imaging. In ptychography, the beam under analysis is scattered off an object, forming a diffraction pattern. Phase retrieval algorithms are used to reconstruct the initial field from the diffraction patterns \cite{BATEY_UM_2014,Goldberg_OL_2023}.

(4) Measurement methods such as RED-SEA-TADPOLE \cite{Gallet_OL_2014} characterize STCs by the spectral interference of the unknown test pulse with a known reference pulse.

(5) Yet another method, STRIPED-FISH \cite{Gabolde_OE_2006}, is based on holography and requires a spatially filtered reference beam. Recently, a similar method, CMISS \cite{Xu_OL_2022}, which does not require a special reference pulse, was proposed.

(6) Finally, there are STC measurement methods based on broadband Young's Double Slit Interferometry. The method relies on the fact that the far-field diffraction pattern of an ultra-short beam that impinges on Young's double slit (or two-pinholes) contains information about the time delay between the two sub-pulses that go through the slits (or pinholes) \cite{Netz_APB_2000,Smartsev_JOPT_2022}. 

While many of these techniques can effectively yield the spatiotemporal characterization of the beam, they generally have relatively complex experimental setups that can be expensive and challenging to install. Moreover, many of these methods rely on optical components that are not typically utilized in the beamline, such as beam splitters, band-pass filters or gratings. As a result, they cannot be employed as \emph{in-situ} diagnostics, and their accuracy in representing the true spatiotemporal field of the experimental focal spot is limited. The most widely used techniques suffer from the need to take dozens or even hundreds of measurements, limiting their usefulness in high-power, low-repetition rate systems. The single-shot techniques, meanwhile, are often experimentally cumbersome. These limitations have made developing next-generation spatiotemporal measurement techniques a hot topic in the world of high-power lasers.

\begin{figure}[t]
            \centering
		\includegraphics[trim={0 30pt 0 0},clip,width=0.70\linewidth]{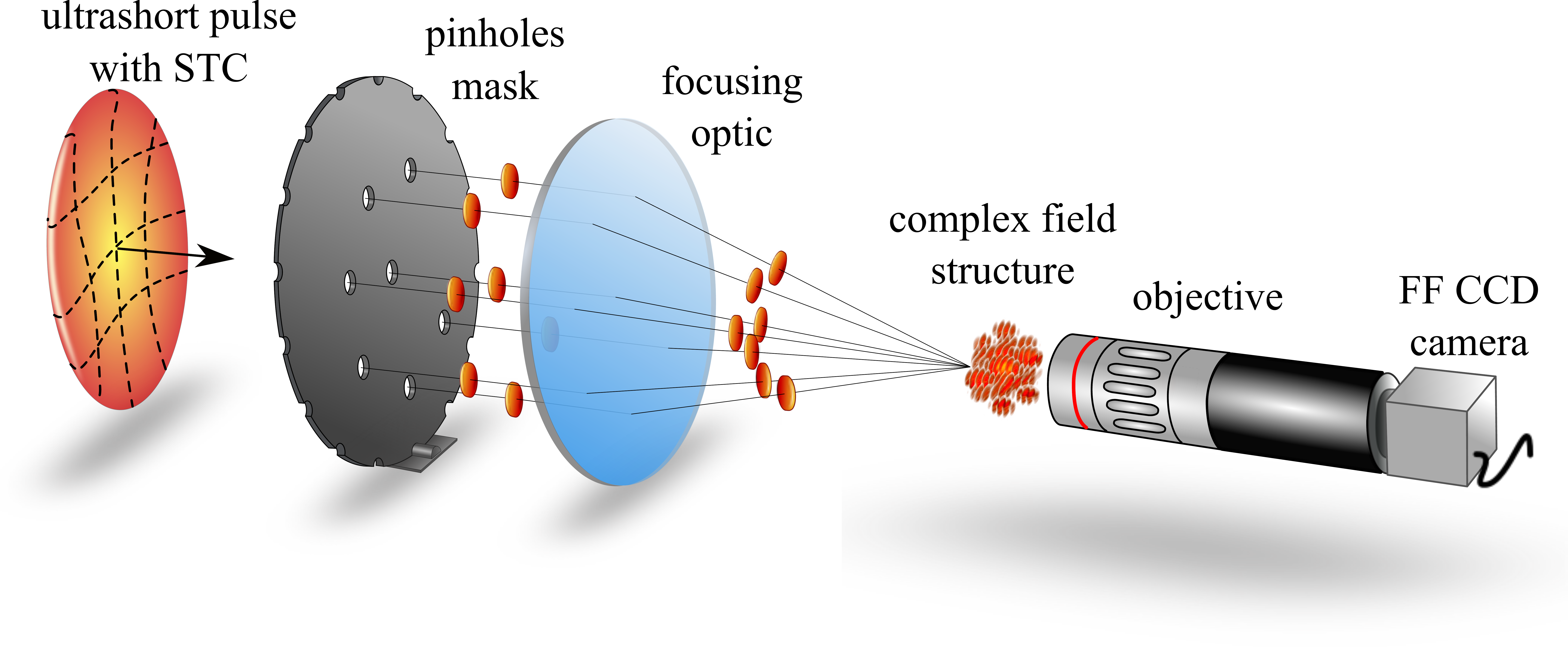}
		\caption{\label{fig:Setup} \small {Simplified experimental setup. A mask splits an ultrashort laser pulse with STCs into dozens of beamlets. A focusing optic concentrates the beamlets at its focal plane, which is then imaged by a microscope objective onto the FF CCD camera, thus registering the polychromatic FF intensity.}}
\end{figure}

This letter introduces an experimentally simple and novel method for measuring the multispectral wavefronts of ultrashort laser pulses in a single shot or only a few shots. IMPALA, or Iterative Multispectral Phase Analysis for LAsers, is based on the linear far-field interferometry of multiple beamlets generated by a special pinhole mask. Instead of a two-pinhole mask, we generalize the Young's Double Slit Interferometry method to dozens of pinholes. The pinholes are arranged in a specific manner to ensure that each interference pattern created by any pair of holes does not overlap with others in the spatial Fourier plane, an idea inspired by phase retrieval in randomly positioned cores of a fiber bundle \cite{Kogan_17}. In one shot, it allows for the retrieval of spectrally resolved wavefronts of the ultrashort beam, spatially sampled at the pinhole positions. This is sufficient for low spatial resolution wavefront retrieval. In addition, rotations of the mask allow for improved spatial resolution of the wavefronts. This particular implementation of the pinhole mask was optimized to get a high spatial resolution from 12 shots, each with a unique rotation of the mask. Our method is remarkably straightforward because the only non-standard optical element it requires is a special pinhole mask placed before the focusing optics used in the experiment. This mask can be easily cut or 3D printed. The method is minimally intrusive and can be readily moved in and out of an existing optical setup. It is easily scalable to different beam sizes, focal lengths of the optics, and spectral bandwidths. By combining scalability, a simple setup, low cost, and single shot compatibility with sufficient spatial and spectral resolution for many applications, IMPALA has the characteristics necessary to become a standard diagnostic tool.

\begin{figure*}[t]
\centering
		\includegraphics[trim={0 1pt 0 0pt},clip,width=1\linewidth]{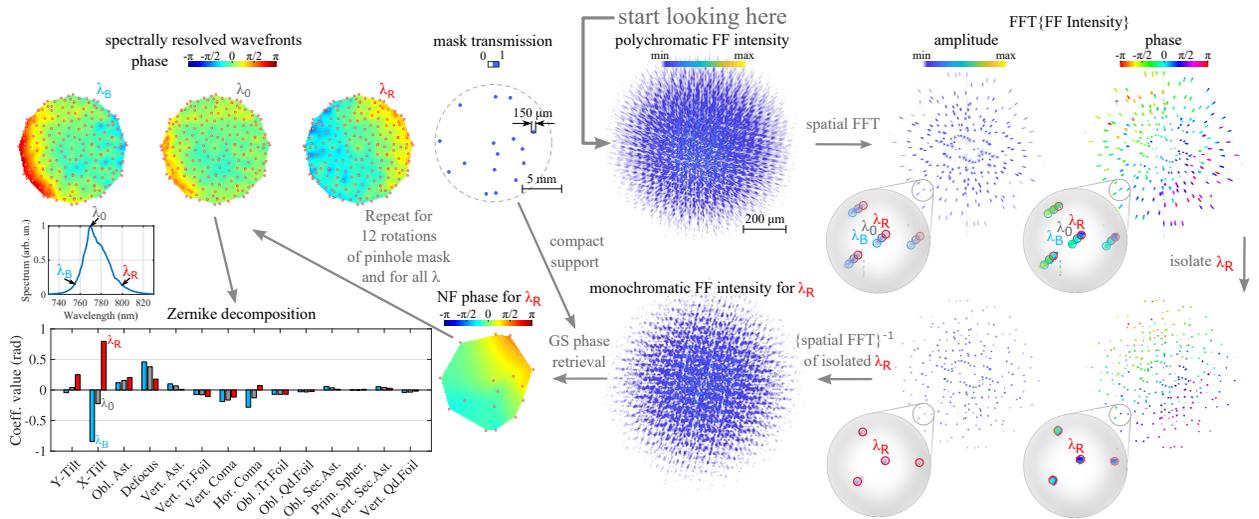}
		\caption{\label{fig:Algo} \small {Algorithm structure and sample wavefront retrieval. The monochromatic FF intensity is extracted from the measured polychromatic FF intensity by utilizing the spatial separation of the colors in the streaks. The GS algorithm reconstructs the NF phase of the beamlets using the mask transmission function (value of 1 at holes, 0 everywhere else) as compact support for the NF amplitude. The process is repeated for different colors and mask rotations to obtain spectrally resolved wavefronts, from which the Zernike coefficients of the analyzed wavefronts are extracted.}}
\end{figure*}

\section{Experimental methods and algorithm}
The simplified experimental setup is depicted in Fig.~\ref{fig:Setup}. An ultrashort laser pulse with unknown STCs impinges on the mask and is focused by a parabolic mirror or lens. A microscope objective images the optic's focal plane onto a CCD camera where the beamlets interfere and form a speckle pattern, shown in Fig.~\ref{fig:Algo} as the polychromatic far-field (FF) intensity map.

As we showed in \cite{Smartsev_JOPT_2022}, each pair of broadband beamlets form a structured fringe pattern at the focus in the FF. The spatial fast Fourier transform (FFT) of this fringe pattern consists of a central peak and two sideband streaks, whose positioning depend on the beamlets' relative orientation. The length of the streaks depends on the bandwidth of the beam and also on the hole separation distance (See Eq. 16 in \cite{Smartsev_JOPT_2022}). These streaks contain temporal information about the relative group delay between the beamlets. This is equivalent to a PFT estimate, which is proportional to the linear term in a spectrally resolved wavefront sampled at two spatial points. Generalizing this method allows for the extraction of the spectrally-resolved wavefront, sampled at a dozen points. We obtain the optimal hole arrangement for a given number of holes and given laser spectrum using a genetic algorithm that packs the streaks with minimum overlap (see Appendix) to ensure proper phase retrieval.

The IMPALA algorithm is schematically shown in Fig.~\ref{fig:Algo}. The method is based on a measurement of the polychromatic FF intensity and utilizes the compact support constraint of the pinhole mask as a substitute for the NF amplitude \cite{Fineup_1987}. The measured FF speckle intensity is Fourier transformed, and the resultant intensity (FFT$\{$FF Intensity$\}$) consists of sparse streaks. Different colors of the beam are distributed radially along these streaks. For example, along the same streak, the "redder" component (in our case $\lambda_R=795$ nm) is located closer to the center when compared to the "bluer" component ($\lambda_B=758$ nm) and the central component ($\lambda_0=772$ nm) is between them. This is due to the fact that the "redder" monochromatic speckle pattern has a smaller spatial frequency, which is inversely proportional to the wavelength. This separation allows us to isolate monochromatic "spots" inside the polychromatic streaks (see Appendix). These isolated spots (and their corresponding phases and amplitudes) inside of the Fourier-transformed speckle intensity are spatially transformed back to obtain a monochromatic speckle pattern. Having obtained a monochromatic intensity in the FF and using the constrained compact support of the pinhole mask to substitute the mask transmission function (value of 1 at the holes, 0 everywhere else) for the NF amplitude, we use the GS algorithm to find the spatial phase (and amplitude: see Appendix) in the NF for each specific color. The process is then repeated for different colors to obtain a spectrally-resolved wavefront. The spatial resolution is enhanced by rotating the mask over 12 angular positions, obtaining more dense spatial sampling in the NF. It should be noted that the hole placement in the mask was optimized to balance between the sparsity of speckle intensity in the Fourier plane (avoid overlapping streaks) and uniform sampling of holes in the NF when summing all 12 rotations (see Appendix).

The length of the polychromatic streak (along the radial coordinate) defines the spectral resolution of the phase measurement. Drawing from the model derived in \cite{Smartsev_JOPT_2022}, we show in Appendix that the spectral resolution is $\Delta \lambda_{impala}\approx \Delta \lambda(1+[\Delta \lambda s_b / \sqrt{2 \log 2} \lambda_0 d_b]^2)^{-1/2}$, where $\Delta \lambda$ and $\lambda_0$ are the full width at half maximum (FWHM) of the spectrum and its central wavelength, while $s_b$ and $d_b$ are the beamlets' separation and diameter (or hole diameter) in the NF, respectively. For the longest streaks in Fig.~\ref{fig:Algo}, the spectral resolution is $\Delta \lambda_{impala}\approx 10$ nm, using the experimentally relevant parameters $\Delta \lambda=30$ nm, $\lambda_0=770$ nm, $s_b=12$ mm, and $d_b=150$ \textmu m. 

It should be noted that IMPALA, as a linear measurement, requires a non-linear measurement of spectral phase in order to stitch together the monochromatic wavefronts.

\section{Results and discussion}
The experiments were performed at Laboratoire d’Optique Appliquée using the Salle Noire 3.0 laser system. The laser's front end consists of a commercial Ti:Sa oscillator (Rainbow, Femtolasers GmbH), followed by a Ti:Sa chirped pulse amplifier system (Femtopower Pro-HE) delivering 30 fs compressed pulses of up to 1.2 mJ energy with a 1 kHz repetition rate. 
We used a standard cube polarizer-waveplate pair attenuator to obtain a variable attenuation of the beam in the range of hundreds of \textmu J without introducing any STC between the measurements. In addition, a neutral density filter with an optical density of 4 was installed before the FF CCD sensor during all measurements. To avoid any risk of non-linear effects, we stretched the pulse to nearly 1 ps by introducing GDD through an acoustic-optic programmable dispersive filter (Dazzler, Fastlite) integrated into the main amplifier. We tested IMPALA by measuring the spectrally resolved wavefronts of the laser. For further testing, we introduced STCs into the laser in a controlled way and retrieved them with IMPALA, thus verifying the correctness of our method.
\subsection{Parabolic mirror as a focusing optic and introduced PFT}
\begin{figure}[t]
\centering
		\includegraphics[trim={0 1pt 0 0pt},clip, width=0.50\linewidth]{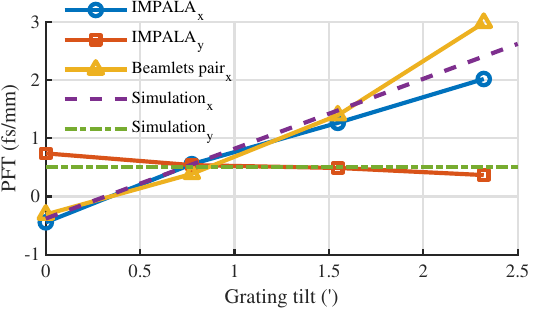}
		\caption{\label{fig:PFT} \small {Measured and simulated introduced PFT as a function of a grating tilt of the compressor.}}
\end{figure}

We tested our method by introducing a controlled amount of PFT by acting on the parallelism of the compressor gratings in their dispersion plane (See Appendix). The focusing element, in this case, was a 100 mm focal length parabolic mirror. As seen in Fig.~\ref{fig:PFT}, we observe a linear PFT as a function of grating misalignment in the x-axis (grating dispersion plane), while the PFT in the y-axis is nearly constant. The slight variation can be explained by an imperfect alignment of the compressor's roof mirrors, which can couple the x and y axes for PFT. In addition, we estimate the PFT with our previous method based on the interference of two beamlets by using a mask with two holes \cite{Smartsev_JOPT_2022}. The IMPALA's PFT measurement agrees both with the measured PFT using the beamlets method and the simulation. The measured wavefronts for a grating tilt of 2.3' (minutes of arc) and their Zernike decomposition are shown in Fig.~\ref{fig:Algo}.

\subsection{Singlet lens as a focusing optic and introduced PFC}
\begin{figure}[t]
\centering
		\includegraphics[width=0.6\linewidth]{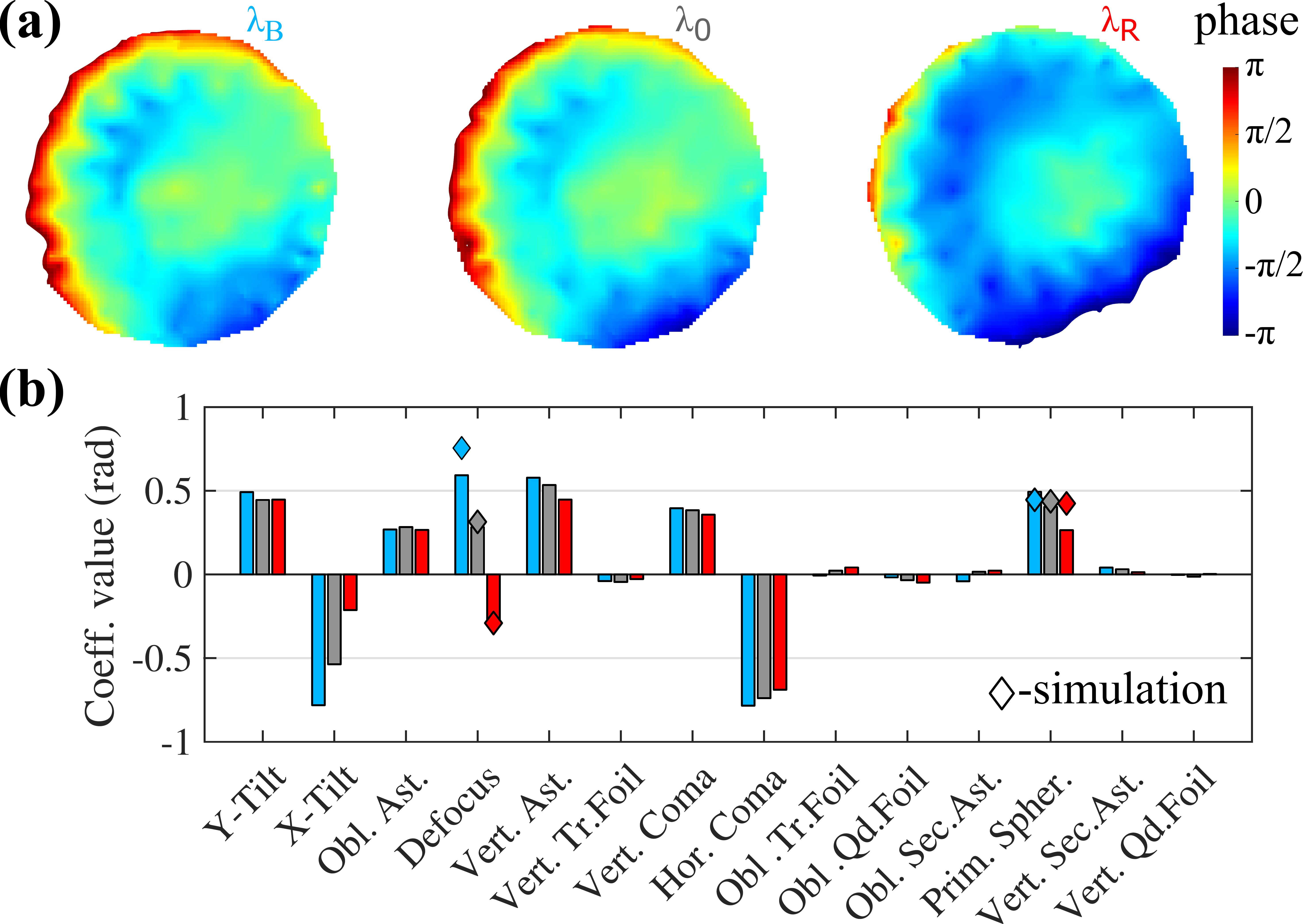} 
		\caption{\label{fig:Singlet} \small {Measured wavefronts (a) and their Zernike decomposition (b) for experiments with a singlet lens as a focusing optic. Predicted chromatic defocus and spherical terms of the lens are shown as diamonds.}}
\end{figure}

We also tested IMPALA by using a spherical singlet lens as a focusing element. A singlet lens introduces PFC when focusing an ultrashort pulse since it has a non-negligible amount of longitudinal chromatic aberration \cite{Bor_89}. We used a 100 mm focal length plano-convex singlet lens as the focusing element (See Appendix). The measured spectrally resolved wavefronts are depicted in Fig.~\ref{fig:Singlet} (a). These spatial phase fronts were unwrapped manually at a few points by adding or removing $2\pi$. The displayed phase range is limited to 2$\pi$ to make the spherical term visible. From the Zernike decomposition shown in Fig.~\ref{fig:Singlet} (b), it is evident that the lens introduces non-negligible chromatic defocus terms, observable in the measurement through the chromatically dependent defocus term. This corresponds to a PFC of: \mbox{PFC$_{meas~singlet}$=0.20 fs/mm$^2$} (while the predicted \mbox{PFC$_{pred~singlet}$=0.23 fs/mm$^2$}). For comparison, the value of the PFC in the measurement with the parabola, shown in Fig.~\ref{fig:Algo}, is \mbox{PFC$_{parab}$=0.052 fs/mm$^2$}. Additionally, the lens introduces the expected spherical aberration, and the measured value quantitatively aligns well with the simulation (diamonds in Fig.~\ref{fig:Singlet}). The other measured aberration terms (astigmatism and coma) are probably mainly due to imperfect lens alignment. The non-zero value of PFC$_{parab}$ is expected and reflects the fact that there is a chromatic beam expander in the laser chain.

\section{Conclusion}
We presented a simple yet powerful method to measure spectrally resolved wavefronts of broad spectrum lasers. We successfully tested our method by inducing controlled STCs and demonstrating their accurate measurement. Our method performed well both when compared to simulations and when benchmarked against a previous measurement method. This method is extremely simple to implement and to scale to different systems and is compatible with high power, low repetition rate laser systems. It thus promises to become a new standard in high-power laser laboratories, potentially making STC measurements as common a part of laser diagnostics as spatial wavefront and temporal measurements are today.

\subsection*{Funding} 
European Union's Horizon 2020, no. 101004730 and Advanced Grant ExCoMet 694596; Agence Nationale de la Recherche, no. ANR-20-CE92-0043-01 and no. ANR-10-LABX-0039-PALM. LaserLAB-Europe no. 871124; Schwartz/Reisman Center for Intense Laser Physics, Benoziyo Endowment Fund for the Advancement of Science, the Israel Science Foundation, Minerva, Wolfson Foundation, the Schilling Foundation, R. Lapon, Dita and Yehuda Bronicki, WIS-CNRS (IPR LAMA).

\subsection*{Acknowledgments}
We thank Prof. Dan Oron for the fruitful discussions. We also thank Dr. Nicolas Thurieau for printing the masks and Sébastien Brun and Pascal Rousseau for assistance in micro-drilling.

\subsection*{Disclosures}
The authors declare no conflicts of interest.

\subsection*{Data Availability Statement}
The data supporting this study's findings are available from the authors upon reasonable request.

\appendix

\renewcommand{\thefigure}{A\arabic{figure}}
\renewcommand{\theequation}{A\arabic{equation}}

\setcounter{figure}{0}

\section*{\underline{Appendix}}

\section{Phase retrieval algorithm}

\begin{figure}
\centering
		\includegraphics[width=1.0\linewidth]{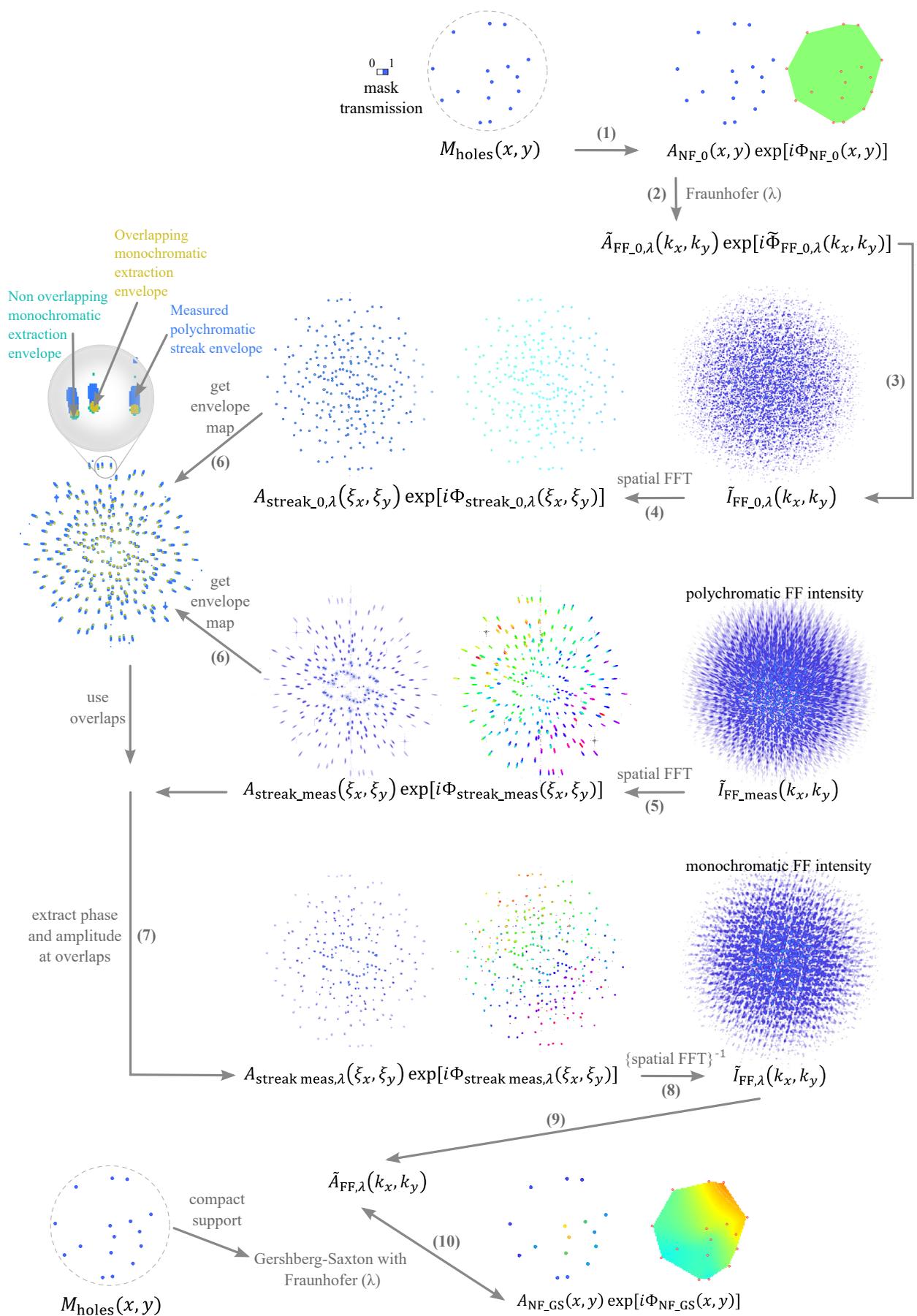}
  \centering
		\caption{\label{fig:algo} \small {Schematic presentation of IMPALA algorithm implementation.}}
\end{figure}

The retrieved wavefronts presented in the main paper were obtained using the IMPALA algorithm, the implementation of which is depicted schematically in Fig.~\ref{fig:algo}. The steps for the phase retrieval of one show (one mask rotation) and one specific wavelength, $\lambda$, are as follows:

\subsection{Extraction of monochromatic speckle pattern from data}
First we extract the monochromatic speckle pattern from the measured polychromatic speckles.
(1) We begin by constructing what we call the pinhole mask transmission function $M_{holes}(x,y)$, (holes shown larger in the figure for clarity), which is zero outside the pinholes and 1 inside the pinholes. We use this to generate a NF (near field) amplitude which has the same amplitude behaviour (zero outside the holes and unity inside the holes), $A_{NF~0}(x,y)$ and has a flat phase $\Phi _{NF~0}(x,y)=0$. We call this the "extraction" NF.

(2) This "extraction" field is propagated to the focal plane, which is the FF (far-field) plane, using the Fraunhofer propagator for the specific wavelength $\lambda$. 

(3) Taking the absolute value squared of $\tilde{A}_{FF~0,\lambda}(k_x,k_y)\exp{[i\tilde{\Phi}_{FF~0,\lambda}(k_x,k_y)]}$ results in the "extraction" FF speckle intensity, $\tilde{I}_{FF~0,\lambda}(k_x,k_y)$.

(4) The spatial FFT (fast-Fourier transform) of this "extraction" monochromatic speckle intensity then reveals the "short monochromatic streaks" structure $A_{streak~0, \lambda}(\xi_x,\xi_y)$ needed for the localization of the places where the amplitude and phase will be extracted from the experimental streaks. It could be noted that the amplitude $A_{streak~0, \lambda}(\xi_x,\xi_y)$ and the phase $\Phi_{streak~0, \lambda}(\xi_x,\xi_y)$ are "flat" since they correspond to the "flat" NF amplitude and phase that we assumed in step (1).

(5) Next, we perform a spatial FFT on the measured polychromatic speckle pattern $\tilde{I}_{FF~meas}(k_x,k_y)$ and get the measured phase $\Phi_{streak~meas}(\xi_x,\xi_y)$ and amplitude $A_{streak~meas}(\xi_x,\xi_y)$. These polychromatic streaks contain the amplitude and phase for the continuous spectrum. 

(6)  Now, we take the amplitude $A_{streak~0, \lambda}(\xi_x,\xi_y)$, obtained in step (4), and turn non-zero values to 1. This is now an "envelope map", which will be used to localize the places where the experimental amplitude and the phase will be extracted from the measured polychromatic streaks for this specific wavelength.

To isolate the single color contribution from the polychromatic streaks, we also generate an "envelope map" from the polychromatic streaks by turning the non-zero values to 1. We now want to maximize the overlap of the two maps, since poor overlap of the maps will result in incorrect amplitude and phase values in the non-overlapped areas. To account for experimental uncertainties, we optimize the overlap by allowing small perturbations of the hole positions (300 \textmu m range for each hole separately) in the NF. This process is done for each of the rotations. It can be seen that the overall overlap of these envelopes is good after the optimization, though some mismatch remains as a source of error.

(7) Using these overlapped maps, we extract the measured phase and amplitude at the overlap locations, yielding a monochromatic streak \, $A_{streak~meas, \lambda}(\xi_x,\xi_y)\exp{[i\Phi_{streak~meas, \lambda}(\xi_x,\xi_y)]}$, from the polychromatic data.

(8) Obtaining the monochromatic streaks with the measured phase and amplitude allows us to get the monochromatic speckles in the FF, $\tilde{I}_{FF,\lambda}(k_x,k_y)$, by applying an inverse spatial FFT.

(9) We get the field amplitude, $\tilde{A}_{FF,\lambda}(k_x,k_y)$, from the intensity and we initially assume a flat phase for the subsequent GS retrieval.

\subsection{Gerchberg-Saxton phase retrieval}

(10) The final step is to retrieve the phase in the NF iteratively using a standard GS (Gerchberg-Saxton) algorithm with compact support constraint from the pinhole mask in the NF \cite{Fineup_1987}. Compact support for the NF means that we allow the algorithm to converge to some NF field bounded by the holes of the mask in the NF, while zeroing the field outside the mask.

The following GS iterations are for a specific $\lambda$ and so we will omit the $\lambda$ subscript and the coordinate notation for clarity:

(GS step 0) start with a specific amplitude and flat phase (or zero phase) from step (9) in FF: $\tilde{A}_{FF}\exp{[i\tilde{\Phi}_{FF,flat}]}$

Iteration 1:

(GS step 1) propagate to NF with Fraunhofer and get: ${A}_{NF, iter1}\exp{[i{\Phi}_{NF,iter1}]}$

(GS step 2) constrain the field by the compact support of the mask: ${A}_{NF, iter1}\exp{[i{\Phi}_{NF,iter1}]}M_{holes}$

(GS step 3) propagate to the FF with Fraunhofer and get: $\tilde{A}_{FF,iter1}\exp{[i\tilde{\Phi}_{FF,iter1}]}$

(GS step 4) replace the FF amplitude with the specific amplitude we started in (GS step 0) and preserve the phase: $\tilde{A}_{FF}\exp{[i\tilde{\Phi}_{FF,iter1}]}$

Iteration 2:

(GS step 1) propagate to NF with Fraunhofer and get: ${A}_{NF, iter2}\exp{[i{\Phi}_{NF,iter2}]}$

and so on...

Propagating between the NF and the FF using the Fraunhofer propagator while preserving the field amplitude in the FF and constraining the NF amplitude not to leak out from the mask localization allows the algorithm to converge to the NF phase and amplitude solution. We are not interested here in the FF phase but in the phase in the NF, $\Phi _{NF~GS}(x,y)$. It should be noted that for each GS iteration, we constrained each hole phase to be a constant value across the hole. The convergence could be monitored by observing the reducing and converging value of the error for the nth iteration comparing $\tilde{A}_{FF, iter~n}$ with  $\tilde{A}_{FF}$. For our specific runs we observed convergence after 25-35 iterations.

The end result of these iterations is a retrieved wavefront $\Phi _{NF~GS}$, and amplitude $A _{NF~GS}$, sampled at the hole positions, at a specific wavelength. Combining additional rotations of the mask and different wavelengths yields the spectrally resolved wavefront of the beam.

\newpage
\section{Retrieved amplitude}
\begin{figure}
\centering
		\includegraphics[width=0.7\linewidth]{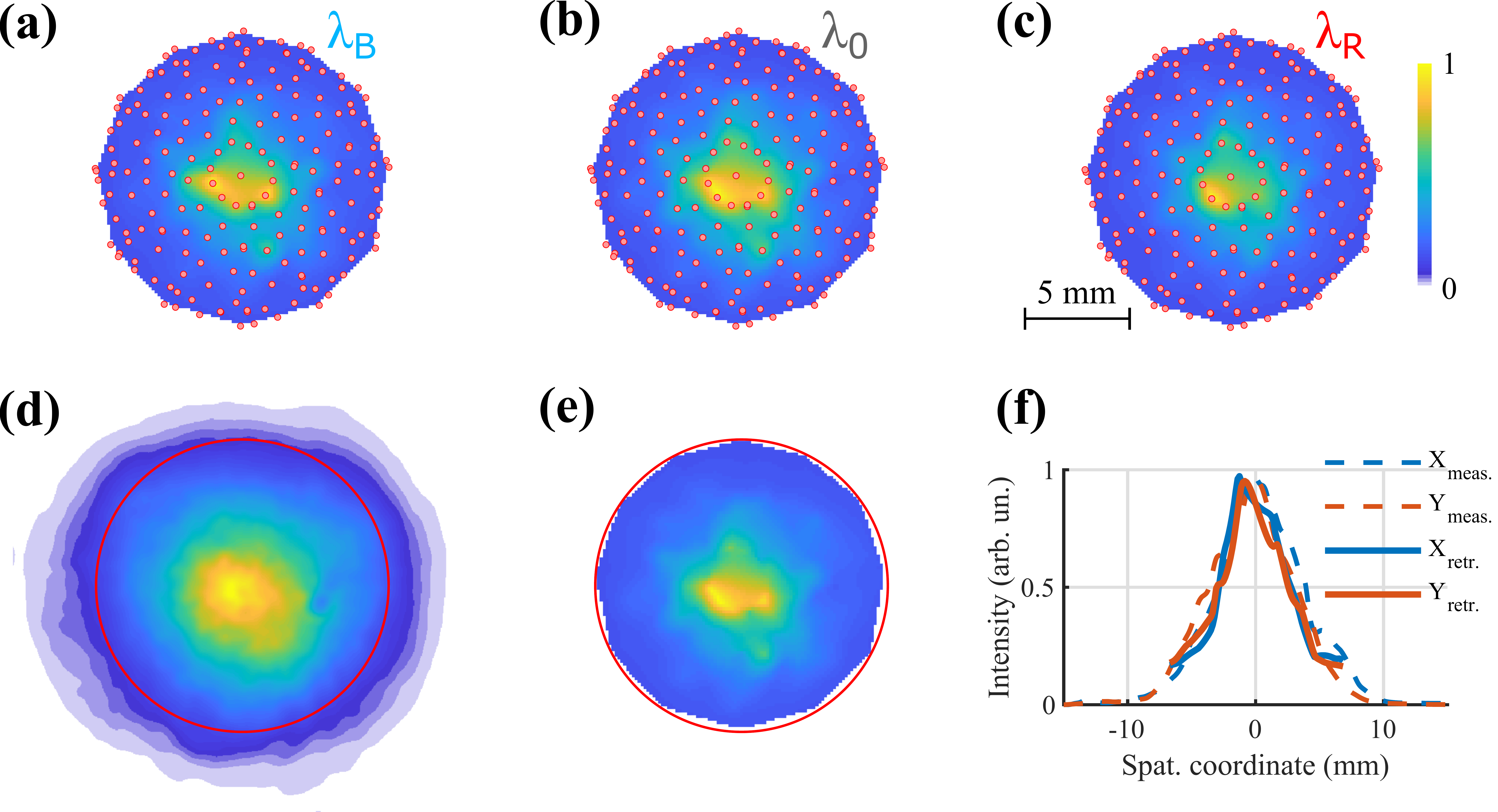}
		\caption{\label{fig:NF_int} \small {(a-c) Reconstructed monochromatic amplitudes (d) Measured NF amplitude from scattered light (e) Reconstructed polychromatic amplitude (f) Comparison between the measured and reconstructed intensity X-Y cross-sections.}}
\end{figure}

By using the compact support of the pinhole mask, in addition to the retrieved phase, the algorithm retrieves the amplitude sampled at the position of the holes as it converges to a solution. The retrieved amplitude provides relative amplitude ratios between the sampled points. We normalize and offset the retrieved intensities (amplitude squared) so that the maximal value is unity and the average edge value (along the maximal measured radius) corresponds to the average edge value of the measured intensity. The retrieved intensities associated with the spectrally resolved wavefront depicted in Fig. 2 in the main paper for the three wavelengths (758, 772, and 795 nm) are depicted in Fig.~\ref{fig:NF_int} (a-c) respectively. 

\begin{figure}
\centering
		\includegraphics[width=0.7\linewidth]{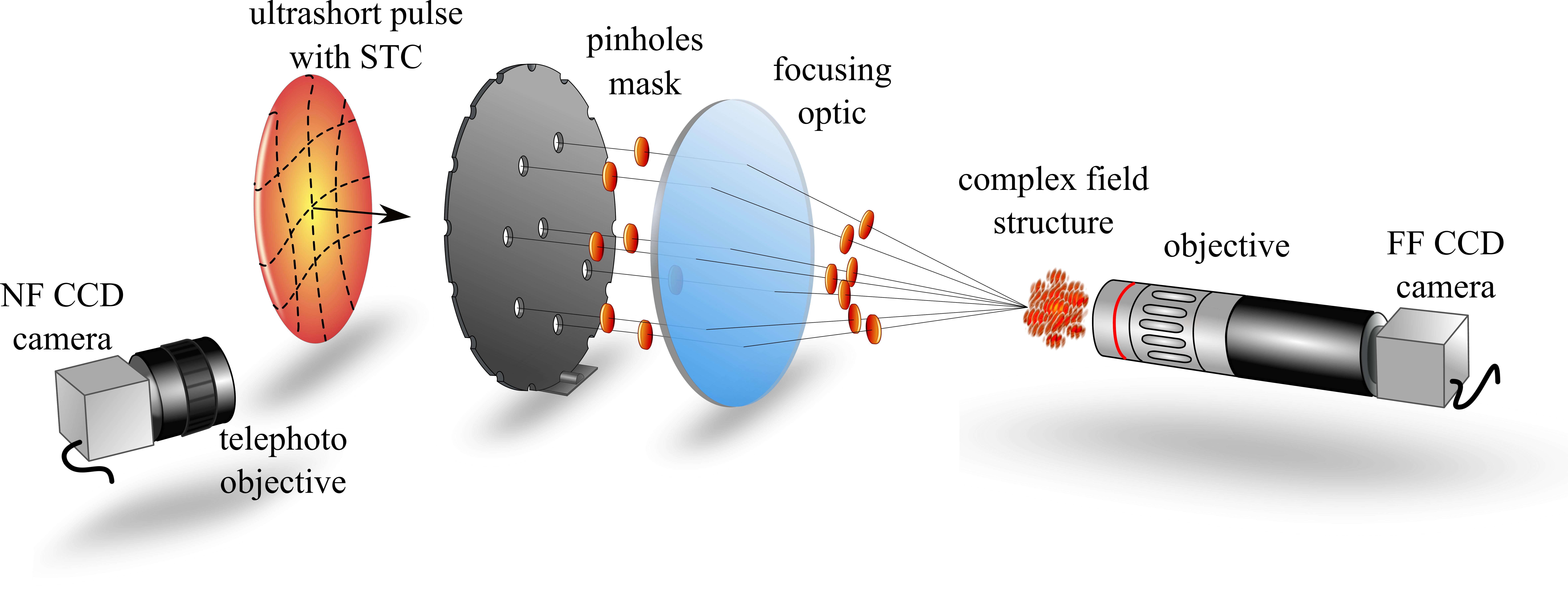}
		\caption{\label{fig:NF_setup} \small {Simplified experimental setup with NF intensity measurement. The NF CCD camera with telephoto objective captures the NF intensity of the beam, which is scattered off a paper attached to the mask.}}
\end{figure}

We measure the NF beam intensity by registering scattered light with a CCD camera and telephoto objective from a paper attached to the back surface of the pinhole mask as shown in Fig.~\ref{fig:NF_setup}. The measured NF intensity profile is depicted in Fig.~\ref{fig:NF_int} (d), while the spectrally weighted sum of the retrieved intensities is in Fig.~\ref{fig:NF_int} (e). The measured area for the retrieved case of 12 mask rotations is marked by a red circle with a 14 mm diameter. The X-Y cross-sections for the measured and retrieved intensities are plotted in Fig.~\ref{fig:NF_int} (f). As we can see the measured and retrieved NF intensity shapes are in reasonable agreement.

\section{Spherical singlet lens wavefronts simulation}
\begin{figure}
\centering
		\includegraphics[width=1\linewidth]{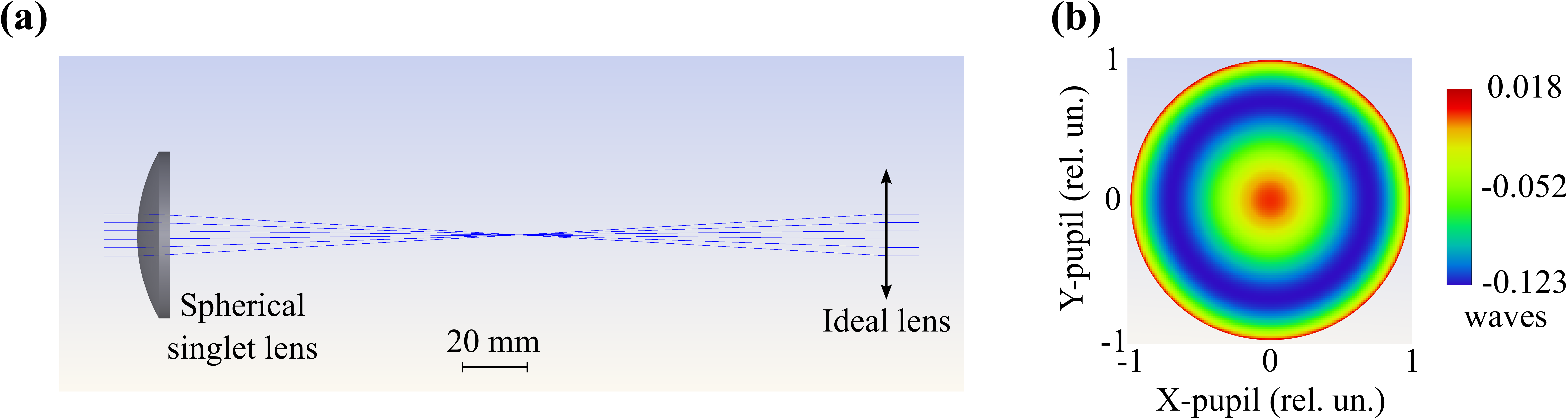}
		\caption{\label{fig:singlet} \small {OpticStudio simulation layout (a) and the simulated wavefront for the 770 nm.}}
\end{figure}

The lens used in the experiment was a 2-inch plano-convex lens CVI-PLCX-50.8-51.5-UV made of fused silica with a radius of 51.5 mm and a central thickness of 10 mm. The approximate focal length of the lens is 100 mm. We simulated the lens-focused spectrally resolved wavefronts and their Zernike decompositions using OpticStudio (Zemax) software. Using its standard tools, we extracted the wavefronts and their Zernike decompositions (shown in the main article, Fig. 4 (b)) after recollimating the beam with an ideal lens. A normalization factor of $2(\pi)^{3/2}$ for Zernike terms extracted from OpticStudio was used to change the units to radians. The optical layout is depicted in Fig.~\ref{fig:singlet} (a). The wavefront for the 770 nm wavelength is shown in Fig.~\ref{fig:singlet} (b).

\section{Compressor induced PFT simulation}
\begin{figure}
\centering
		\includegraphics[width=0.9\linewidth]{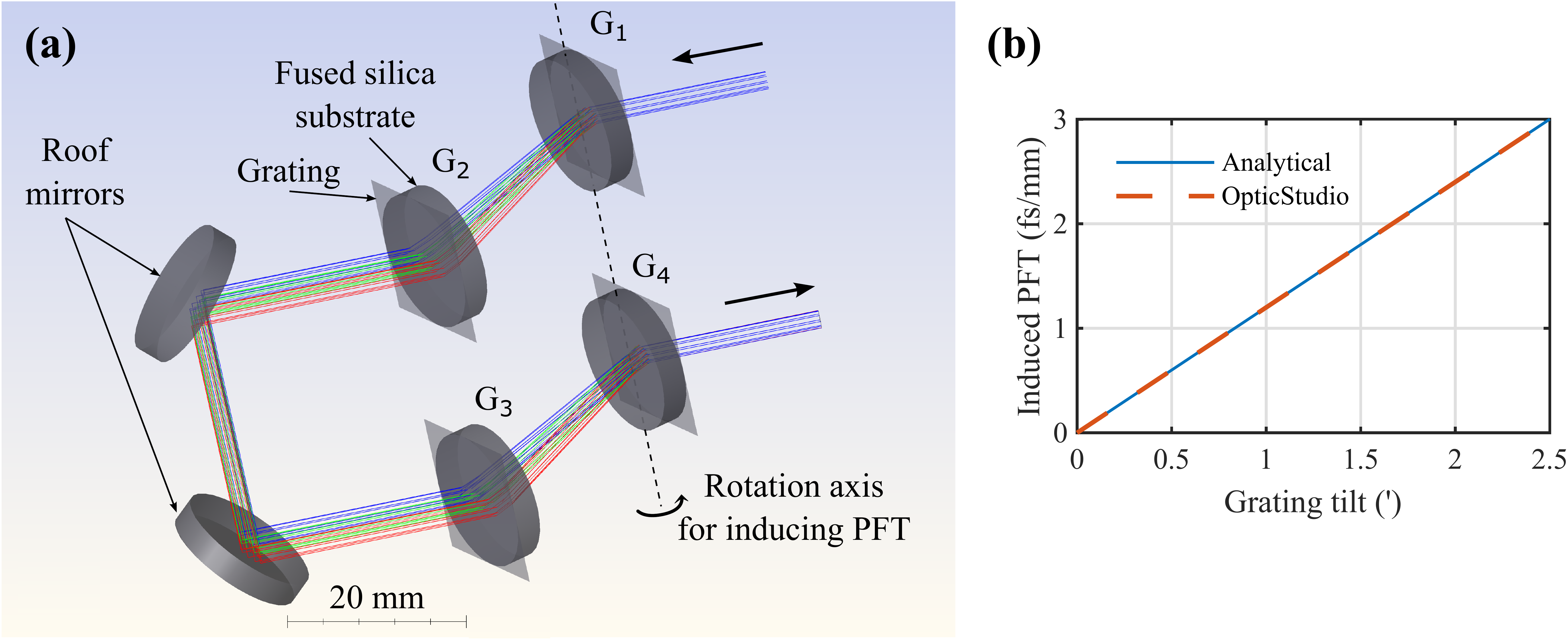}
		\caption{\label{fig:comp} \small {Optical layout of the compressor used in the simulation (a) and induced PFT as a function of the grating misalignment angle in minutes of arc (') (b).}}
\end{figure}
We simulated the transmission grating compressor used in the experiment with the OpticStudio software and computed the induced PFT as a function of the grating misalignment angle. The optical layout is depicted in Fig.~\ref{fig:comp} (a). The compressor consists of a pair of transmission gratings which each have a 3 mm fused silica substrate and an engraved grating with a density of 1280 lines/mm. The grating surface is shown as a thin rectangular surface, and the fused silica substrate is shown as a cylindrical volume. The nominal grating angle to the incoming beam in the dispersion plane is 31 degrees. The compressor was simulated in sequential mode, which simulates a single pass through 4 gratings: G$_1$, G$_2$, G$_3$, G$_4$, rather than a double pass through 2 gratings as we had in the experiment, where G$_1$ and G$_4$ are the same optical elements, as are G$_2$ and G$_3$. To simulate induced PFT, we perturbed the angle of rotation of the first grating along the axis shown in Fig.~\ref{fig:comp} (a). For each angle, we calculated the induced amount of PFT with a special macro \cite{Slava_Thesis}, which directly extracts the value of the group delay of the pulse along the transverse spatial coordinate. The simulated PFT for small angles is in good agreement with a simple analytical estimation, as presented in Fig.~\ref{fig:comp} (b).

The analytical expression for the PFT as a function of the gratings misalignment is based on Eq. 8 in \cite{Zhao_2023}.

\begin{equation}
{\text{PFT}}=2\epsilon N \frac{\lambda_0 \tan{\beta_0}}{c \cos{\alpha}}
\end{equation}

where $\epsilon$ is the misalignment angle (in radians) in the dispersion plane, $N$ is the grating density (1280 lines/mm), $\lambda_0$ is the central wavelength (770 nm), $\alpha$ is the grating angle with respect to the beam (31 degrees), and $\beta_0$ is the first order diffraction angle, such that $\beta_0=\arcsin{(\sin{\alpha}-\lambda_0 N)}$, and $c$ is the speed of light. For example, when the misalignment angle $\epsilon=2.5$ (') (minutes of arc), the resulting PFT$\approx 3$ (fs/mm).

\section{Spectral resolution}
As we have shown in our previous work \cite{Smartsev_JOPT_2022}, for the case of two Gaussian beamlets, the resulting pair of streaks in the Fourier plane of the speckle intensity has a Gaussian form $\propto \exp{a_{SB1}}+\exp{a_{SB2}}$ where the arguments in the exponents are:

\begin{equation} \label{eq1}
a_{SB1,2}=-\frac{[\Delta \omega\tau_1 \sigma_0 \omega_0]^2+(cfk_x\pm \omega_0 s_b)^2+i\tau_1[s_b\Delta \omega^2 c f k_x\mp 4\sigma_0^2 \omega_0^3]}{4\sigma_0^2\omega_0^2+\Delta \omega^2 s_b^2}
\end{equation}
Here, $\Delta \omega$ is the laser bandwidth, $\omega_0$ is the central frequency, $\tau_1$ is the relative delay between the two beamlets, $\sigma_0$ is the spatial width of the beamlets, $ k=\omega/c$ is the k-vector, $f$ is the focal distance of the focusing optics, $s_b$ is the spatial separation of the beamlets in the near-field, and $x$ is the far-field coordinate.

Therefore, the streak's length (in the appropriate units in the spatial Fourier plane) is:
\begin{equation} \label{eq2}
\sigma_{streak}^2=2\sigma_0^2\ \omega_0^2+(\Delta \omega^2\ s_b^2)/2
\end{equation}

The width of the monochromatic streak is $\sigma_{mono~streak}^2=2\sigma_0^2\omega_0^2$ and therefore the minimal relative spectral band is:

\begin{equation}
\Delta \omega _{min}=\Delta \omega \frac{\sigma_{mono~streak}}{\sigma_{streak}}
\end{equation}

For expressing the spectral resolution of IMPALA, it is more convenient to work in units of wavelength. Thus, rather than $\Delta \omega$, the laser bandwidth in frequency, we will use spectral bandwidth, $\Delta \lambda_a$, the bandwidth in units of wavelength. The ratio term of the streak lengths is unaffected since it is a unitless ratio. Thus, the minimal wavelength segment resolved by IMPALA is:

\begin{equation}
    \Delta \lambda_{min}=\Delta \lambda_a\sqrt{\frac{1}{1+(\frac{\Delta \lambda_a s_b}{2\sigma_0\lambda_0})^2}}
\end{equation}

\noindent where $\lambda_0$ is the central wavelength. We want to express this formula in simple experimental parameters. The spectral bandwidth, $\Delta \lambda_a$, is defined for the field amplitude, and since we measure the intensity, there is a factor of $\sqrt2$ in the width that must be added. Also, the full width at half maximum is defined by FWHM$=2\sqrt{\log2}\sigma$, and therefore the FWHM of the intensity spectrum is $\Delta\lambda= (2\sqrt{2\log2}\Delta \lambda_a)/\sqrt2$. The beamlet diameter for the intensity is defined as $d_b=2\sqrt2\sigma_0$. Finally the spectral resolution is:

\begin{equation}
\Delta\lambda_{impala}\cong\Delta\lambda\left(1+\left[\frac{\Delta\lambda s_b}{\lambda_0 d_b\sqrt{2\log 2}}\right]^2\right)^{-1/2}
\end{equation}

\begin{figure}
\centering
		\includegraphics[width=0.9\linewidth]{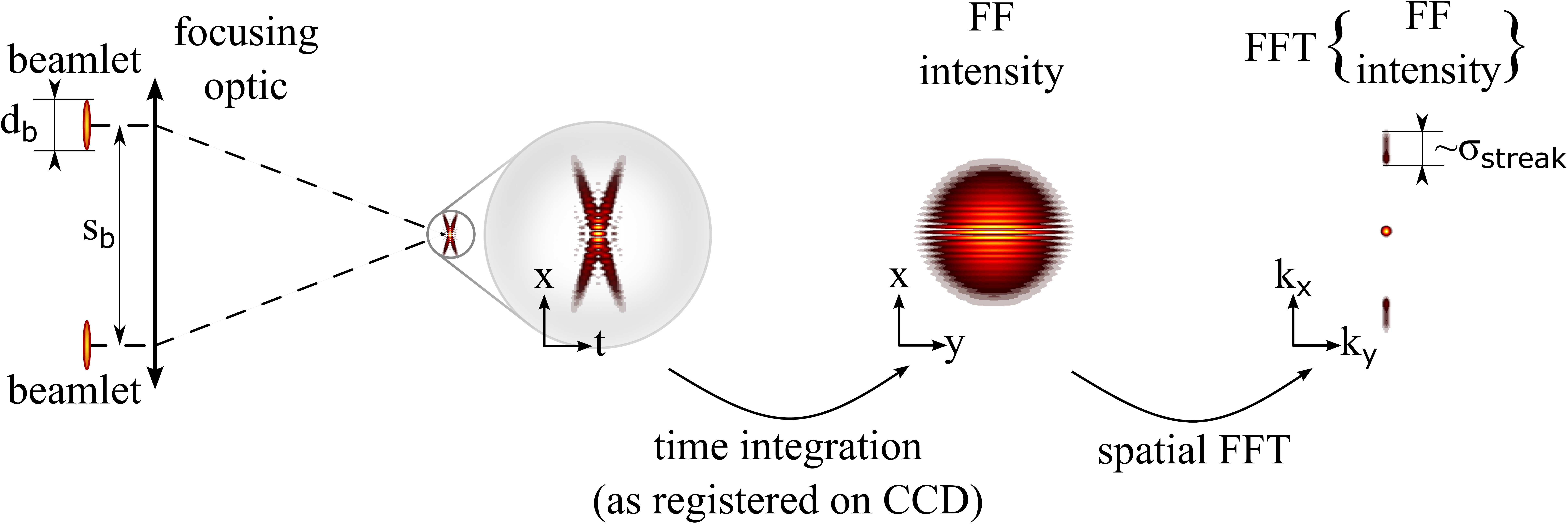}
		\caption{\label{fig:beamlets_geom} \small {Illustration of a pair of beamlets interference with main geometrical parameters and their relevant notation.}}
\end{figure}

Here, $\Delta\lambda$ is the intensity spectral bandwidth FWHM expressed in wavelength. The geometry parameters are shown in Fig.~\ref{fig:beamlets_geom}. For the experimental case presented in the paper, the maximal spectral resolution for the most separated beamlets is $\Delta\lambda_{impala}\cong30~\text{nm}\ast0.3=10~\text{nm}$, and the relevant parameters are $\Delta \lambda=30$ nm, $\lambda_0=770$ nm, $s_b=12$ mm, and $d_b=150$ \textmu m.

\section{Pinhole mask optimization}

The hole arrangement of the mask is optimized to balance between a sparse streak distribution in the Fourier transformed FF and a uniform sampling of holes in the NF when summing all rotations. The optimization procedure is based on the analytical formula for the streak length and width in the Fourier plane of the FF intensity. Each hole pair in the NF generates a symmetrical pattern in the FF intensity Fourier plane. The pattern consists of a central lobe and two streaks. For each hole pair, the length of these streaks is given by $\sigma_{streak}$ from Eq.~\ref{eq2} and depends on the distance of the holes $s_b$, while the streaks' width is $\sigma_{streak}$ at $s_b=0$. For a given number of holes $N_{holes}$, the Fourier-transformed FF intensity will contain $N_{streak}=N_{holes}(N_{holes}-1)/2$ streak patterns. We find each streak's spatial envelope, set its value to 1, and sum all the streak contributions at each location for a given hole distribution. Thus, if two or more streaks overlap, the value in the area of the overlap will be higher than 1. Finally, we integrate the overlapped values,  and normalized by the overlap area for the merit function.
The merit function for the spatial sampling for all rotations is constructed similarly. For each hole in the NF, we set a finite circular area and value of 1, and all holes are summed over all rotations. The overlapped areas which have values higher than 1 are integrated and normalized by the total area. Finally, the algorithm finds the best hole distribution by minimizing both merits simultaneously.

\begin{figure}
\centering
		\includegraphics[width=0.9\linewidth]{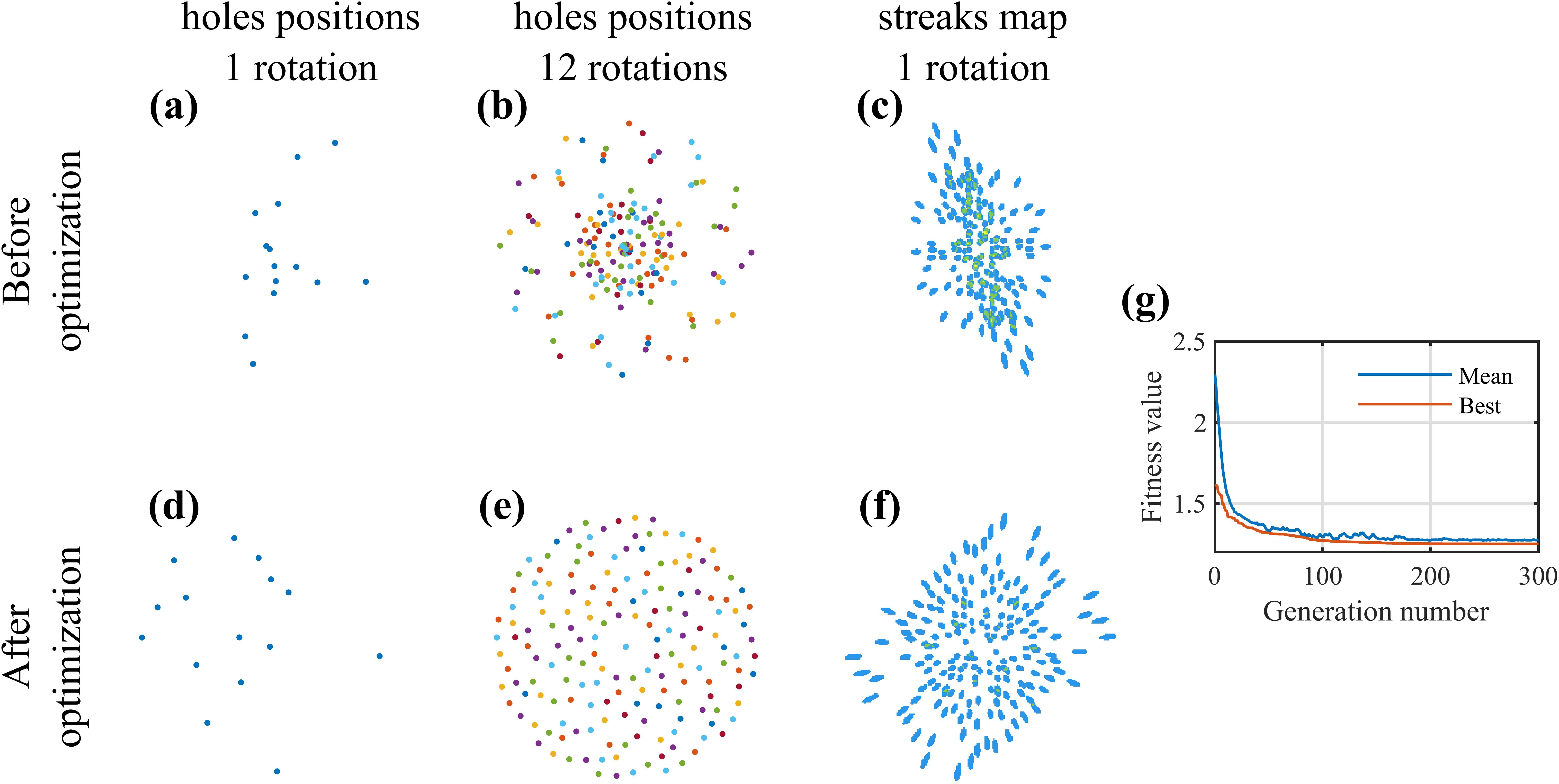}
		\caption{\label{fig:gen_opt} \small {Holes distribution in NF and corresponding streak's maps before and after optimization. (a, d) holes' positions at one rotation, (b, e) holes' positions at 12 rotations, (c, f) streaks maps at one rotation. GA optimization: (g) the fitness of the best individual and the mean fitness values across the entire population as a function of generation number.}}
\end{figure}

We use a genetic algorithm to find the best hole arrangement for a given hole number and given laser parameters. An example of such an optimization for the laser parameters used in the experiment and for 15 holes is shown in Fig.~\ref{fig:gen_opt}. It involves a different arrangement of holes than what was used in the experiment to prepare the visuals in Fig.~\ref{fig:gen_opt}. One hole is constrained to be on the center of the mask. We start from the random hole arrangement depicted in Fig.~\ref{fig:beamlets_geom} (a), and, after 12 rotations, the holes are non-uniformly sampled (b). Prior to optimization, the streaks in the FF Fourier transform have multiple overlaps (c). After the optimization, the hole arrangement (d) brings more uniform sampling after 12 rotations (e) and much less overlap of the streaks (f).


\section{Pinhole mask}

\begin{figure}
\centering
		\includegraphics[width=0.4\linewidth]{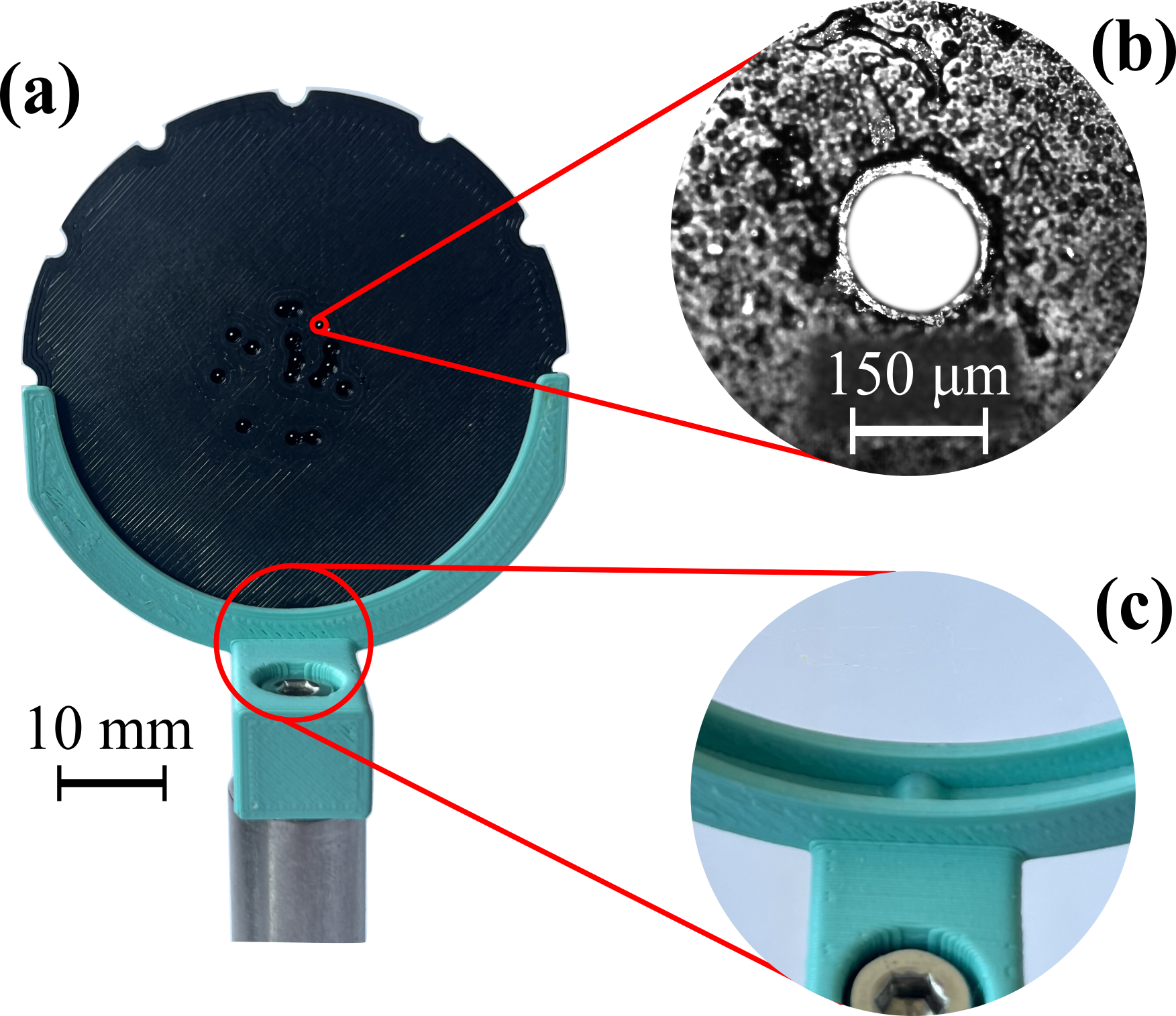}
		\caption{\label{fig:mask} \small {Pinhole mask and its holder used in the experiment (a). Zoomed view of one of the holes (b). A special bump in the mask holder allows for the precise positioning of the mask (c).}}
\end{figure}

The mask holder and mask substrate were 3D printed using fused deposition modeling (FDM) with Polylactic Acid (PLA) and shown in Fig.~\ref{fig:mask} (a). The holder has a special bump shown in a zoomed area in Fig.~\ref{fig:mask} (c) that snaps to the notches made at the edges of the mask. This allows for the precise positioning of the mask for a discrete set of angles. The pinholes of diameter 150 \textmu m were micro-drilled on black aluminum foil (THORLABS-BKF12) (Fig.~\ref{fig:mask} (b)) attached to the mask substrate.

\hspace{1cm}

\bibliographystyle{ieeetr} 
\bibliography{impala}

\end{document}